\documentclass[11pt]{article}
\usepackage{amsmath}
\usepackage{graphicx}
\usepackage{amsfonts}
\usepackage{amssymb}
\usepackage{amscd}
\newcommand{\bc}{\begin{center}}
\newcommand{\ec}{\end{center}}
\newcommand{\be}{\begin{equation}}
\newcommand{\ee}{\end{equation}}
\newcommand{\bea}{\begin{eqnarray}}
\newcommand{\eea}{\end{eqnarray}}
\newcommand{\ba}{\begin{array}}
\newcommand{\ea}{\end{array}}
\newcommand{\edc}{\end{document}}

\def\f{\varphi}

\def\g{\gamma}

\def\O{\Omega}

\def\t{\theta}
\def\b{\beta}
\def\G{\Gamma}

\def\s{\sigma}
\def\m{\mu}

\def\l{\lambda}
\def\L{\Lambda}

\def\d{\partial}

\begin{document}
\thispagestyle{empty}
\begin{center}

{\Large {\bf A Contour Method on  Cayley tree}}
\footnote{The work supported by NATO Reintegration Grant: FEL.RIG.980771.}\\
\vspace{0.4cm}
{\bf U.A. Rozikov}\\
 Institute of Mathematics, 29, F.Hodjaev str., 100125, Tashkent,
 Uzbekistan. E-mail: rozikovu@yandex.ru\\[2mm]

{\it Dedicated to N.N.Ganikhodjaev on the occasion of his 60th
birthday}
\end{center}
\vspace{0.5cm}

{\bf Abstract:} We consider a finite range lattice models on Cayley
tree with two basic properties: the existence of only a finite
number of ground states and with Peierls type condition.  We define
notion of a contour for the model on the Cayley tree. By a contour
argument we show the existence of $s$  different (where $s$ is the
number of ground states) Gibbs measures.

\section{Introduction}

    This paper is a continuation of our previous papers [19]-[21]
devoted to the introduction of a contour method on Cayley tree
(Bethe lattice [1]). The lattice spin systems are large class of
systems considered in statistical mechanics. Some of them have a
real physical meaning, others are studied as suitable simplified
models of more complicated systems.

One of the key problems related to lattice spin systems is the
description of the set of Gibbs measures. The structure of the
lattice plays an important role in the investigations of spin
systems. For example in order to study the phase transition problem
(non-uniqueness of Gibbs measure) for a system on $Z^d$ and on
Cayley tree, respectively, there are two different methods: contour
method (Pirogov-Sinai theory) on $Z^d$ (see e.g.
[2],[5],[7],[14],[16],[17],[22],[24],[25]) and Markov random field
theory on Cayley tree (see e.g. [3],[4],[8-10], [13], [18],[23]).

 In the Pirogov-Sinai theory configurations can be
described by contours which satisfy Peierls condition. This theory
provides tools for a very detailed knowledge of the structure of
Gibbs measures in a region in the relevant parameters space (see
e.g. [22]). The Pirogov-Sinai theory is a low temperature expansion
which enables to control the entropic fluctuations from the ground
states, its natural setup being the lattice systems. But the theory
is not limited to such cases and it has been applied to a great
variety of situations, covering various types of phase transitions.
(see e.g. [6] for details).

Note, that Pirogov-Sinai theory on Cayley tree is not simply
applicable and not much work has been done to develop contour
methods on trees ([19]-[21]). While studying models with the
interaction radius $r\geq 2$ on Cayley tree to describe of Gibbs
measures by the (above mentioned) method of Markov random field
theory becomes difficult, since in this case there appears a set of
nonlinear equations which can not solved analytically. To avoid this
problem it looks very useful to develop a contour method
(Pirogov-Sinai theory) on Cayley tree.

 This paper presents a contour method
for a general model with the interaction radius $r$ ($1\leq r<
\infty$) and with a finite number of ground states on the Cayley
tree of order $k\geq 2.$ For $k=1$ this method was developed in [19]
for a model with nonhomogeneous nearest-neighbor interactions.

The paper is organized as follows. In section 2 we give all
necessary definitions (Cayley tree, model, Gibbs measure etc). In
section 3 under some assumptions on the model (Assumptions A1-A3) we
prove the Peierls condition. Section 4 devoted to definition and
properties of contours on Cayley tree. In section 5 by a contour
argument we show the existence of $s$  different (where $s$ is the
number of ground states) Gibbs measures for the model under
consideration on the Cayley tree of order $k\geq 2$. In the last
section we check our assumptions A1-A3 for several examples of
models.

\section{Definitions}

{\it 2.1. The Cayley tree.} The Cayley tree $\Gamma^k$ (See [1]) of
order $ k\geq 1 $ is an infinite tree, i.e., a graph without cycles,
from each vertex of which exactly $ k+1 $ edges issue. Let
$\Gamma^k=(V, L, i)$ , where $V$ is the set of vertices of $
\Gamma^k$, $L$ is the set of edges of $ \Gamma^k$ and $i$ is the
incidence function associating each edge $l\in L$ with its endpoints
$x,y\in V$. If $i(l)=\{x,y\}$, then $x$ and $y$ are called {\it
nearest neighboring vertices}, and we write $l=<x,y>$.

 The distance $d(x,y), x,y\in V$
on the Cayley tree is defined by the formula
$$
d(x,y)=\min\{ d | \exists x=x_0,x_1,...,x_{d-1},x_d=y\in V \ \
\mbox{such that}  \ \
 <x_0,x_1>,...,<x_{d-1},x_d> \}.$$

For the fixed $x^0\in V$ we set $ W_n=\{x\in V\ \ |\ \
d(x,x^0)=n\},$
$$ V_n=\{x\in V\ \ | \ \  d(x,x^0)\leq n\}, \ \
L_n=\{l=<x,y>\in L \ \ |\ \  x,y\in V_n\}. $$
 Denote $|x|=d(x,x^0)$, $x\in V$.

It is known (see e.g. [8]) that there exists a one-to-one
correspondence between the set  $V$ of vertices  of the Cayley tree
of order $k\geq 1$ and the group $G_{k}$ of the free products of
$k+1$ cyclic groups $\{e, a_i\}$, $i=1,...,k+1$ of the second order
(i.e. $a^2_i=e$, $a^{-1}_i=a_i$) with generators $a_1, a_2,...,
a_{k+1}$.

Let us define a graph structure on  $G_{k}$ as follows. Vertices
which correspond to the "words" $g,h\in G_{k}$ are called nearest
neighbors if either $g=ha_i$ or $h=ga_j$ for some $i$ or $j$. The
graph thus defined is a Cayley tree of order $k$.

For $g_0\in G_k$  a   left (resp. right) transformation shift on
$G_{k}$ is defined by
$$
F_{g_0}h=g_0h \ \ (\textrm{resp.}\ \  F_{g_0}h=hg_0,) \ \  \forall
h\in G_{k}.
$$

It is easy  to see that  the set  of all left  (resp. right) shifts
on $G_{k}$ is  isomorphic to  $G_{k}$.

{\it 2.2. Configuration space and the model.}  For $A\subseteq V$ a
spin {\it configuration} $\s_A$ on $A$ is  defined as a function
 $x\in A\to\s_A(x)\in\Phi=\{1,2,...,q\}$; the set of all configurations coincides with
$\Omega_A=\Phi^{A}$. We denote $\O=\O_V$ and $\s=\s_V.$ Also we
define a {\it periodic configuration} as a configuration $\s\in \O$
which is invariant under a subgroup of shifts $G^*_k\subset G_k$ of
finite index.

More precisely, a configuration $\s\in \O$ is called
$G^*_k-$periodic if $\s(F_yx)=\s(x)$ for any $x\in G_k$ and $y\in
G^*_k.$

 For a given periodic configuration  the index of the subgroup is
called the {\it period of the configuration}. A configuration
 that is invariant with respect to all
shifts is called {\it translational-invariant}.

The energy of the configuration $\s\in \O$ is given by the formal
Hamiltonian
$$
H(\s)=\sum\limits_{A\subset V:\atop {\rm diam}(A)\leq r}I(\s_A)
\eqno (2.1)
$$
where $r\in N=\{1,2,...\},$ ${\rm diam}(A)=\max_{x,y\in A}d(x,y)$,
$I(\s_A): \O_A\to R$ is a given translation invariant potential i.e.
$I(\s_A)=I(\s_{F_yA})$ for any $y\in G_k$. Here
$\s_{F_yA}=\{\s(F_yx), x\in A\}.$

Fix $r\in N$ and put $r'=[{r+1\over 2}],$ where $[a]$  is the
integer part of $a.$ Denote by $M_r$ the set of all balls
$b_r(x)=\{y\in V: d(x,y)\leq r'\}$ with radius $r'$ i.e.
$$M_r=\{b_r(x): x\in V\}.$$

For $A\subset V$ with ${\rm diam}(A)\leq r$ denote
$$n(A)=|\{b\in M_r: A\subset b\}|,$$
where $|A|$ stands for the number of elements of a set $A$.

The Hamiltonian (2.1) can be written as
$$ H(\s)=\sum_{b\in M_r}U(\s_b), \eqno (2.2)$$
where $U(\s_b)=\sum_{A\subset b}{I(\s_A)\over n(A)}.$

For a finite domain $D\subset V$ with the boundary condition
$\varphi_{D^c}$ given on its complement $D^c=V\setminus D,$ the
conditional Hamiltonian is
$$H(\s_D\big
| \varphi_{D^c})=\sum_{b\in M_r:\atop b\cap D\ne \emptyset}U(\s_b),
\eqno(2.3)$$ where
$$ \s_b(x)=\left\{\begin{array}{ll}
\s(x) & \textrm{if \ \
$ x\in b\cap D$}\\
\varphi(x)& \textrm{if \ \ $x\in b\cap D^c.$}\\
\end{array}\right.
$$

{\it 2.3. The ground state.} A {\it ground state} of (2.2) is a
configuration $\varphi$ in $\Gamma^k$ whose energy cannot be lowered
by changing $\varphi$ in some local region. We assume that (2.2) has
a finite number of translation-periodic (i.e. invariant under the
action of some subgroup of $G_k$ of finite index) ground states. By
a standard trick of partitioning the tree into disjoint sets $Q(x)$
centered at $x\in G^*_k$ (the corresponding subgroup of finite
index) and enlarging the spin space from $\Phi$ to $\Phi^Q$ one can
transform the model above into a model with only
translation-invariant or non periodic ground states. Such a
transformation was  considered in [12] for models on $Z^d.$ Hence,
without loss of generality, we assume translation-invariance instead
of translational-periodic and we permute the spin so that the set of
ground states of the model be $GS=GS(H)=\{\s^{(i)}, i=1,2,..,s\},
1\leq s\leq q$ with $\s^{(i)}(x)=i$ for any $x\in V.$

{\it 2.4. Gibbs measure.} We consider a standard sigma-algebra
${\cal B}$ of subsets of $\O$ generated by cylinder subsets; all
probability measures are considered on $(\O,{\cal B})$. A
probability measure $\mu$ is called a {\it Gibbs measure} (with
Hamiltonian $H$) if it satisfies the DLR equation: $\forall$
$n=1,2,\ldots$ and $\sigma_n\in\O_{V_n}$:
$$\mu\left(\left\{\sigma\in\O:\;
\sigma\big|_{V_n}=\sigma_n\right\}\right)= \int_{\O}\mu ({\rm
d}\omega)\nu^{V_n}_\varphi (\sigma_n),\eqno (2.4)$$ where
$\nu^{V_n}_{\varphi}$ is the conditional probability:
$$ \nu^{V_n}_{\varphi}(\sigma_n)=\frac{1}{Z_{n,\varphi}}\exp\;\left(-\beta H
\left(\sigma_n\big |\,\varphi_{V^c_n}\right)\right). \eqno (2.5)$$
Here $\beta={1\over T}, T>0-$ temperature and $Z_{n, \varphi}$
stands for the partition function in $V_n$, with the boundary
condition $\varphi$:
$$Z_{n, \varphi}=
\sum_{{\widetilde\sigma}_n\in\O_{V_n}} \exp\;\left(-\beta H
\left({\widetilde\sigma}_n\,\big
|\,\varphi_{V^c_n}\right)\right).\eqno (2.6)$$

\section{The Peierls condition}

Denote by ${\mathbf U}$ the set of all possible values of $U(\s_b)$
for any configuration $\s_b,$ $b\in M_r.$  Since $r<+\infty$ we have
$|{\mathbf U}|<+\infty.$ Put $U^{\min}=\min \{U: U\in{\mathbf U}\}$
and
$$\l_0=\min\bigg\{{\mathbf U}\setminus \{U\in {\mathbf U}: U=U^{\min}\}\bigg\}-
U^{\min}.\eqno(3.1)$$

The important assumptions of this paper are the following:

{\it Assumption A1.} The set of all ground states is $GS=\{\s^{(i)},
i=1,2,...,s\}, 1\leq s\leq q.$

{\it Assumption A2.} $\lambda_0>0.$

{\it Assumption A3.} Each $\varphi\in GS$ satisfies
$$ U(\varphi_b)=U^{\min} \ \ \mbox{for every}\ \ b\in M_r.\eqno(3.2)$$

{\it Remark.} If a configuration $\s$ satisfies (3.2) i.e.
$U(\s_b)=U^{\min}$ for $\forall b\in M_r$ then it is a ground state.
Moreover for Hamiltonians on $Z^d$ it is well known that a
configuration is a ground state if and only if the condition (3.2)
is satisfied (see e.g. [22]). But such a fact is not clear for
Hamiltonians on the Cayley tree, since the tree is a non-amenable
graph i.e. $\inf\{{|{\rm boundary \, of}\,  W|\over |W|}: W\subset
V, 0<|W|<\infty\}>0$ for $k\geq 2$ (see e.g. [1], [11]).

The {\it relative Hamiltonian} is defined by
$$ H(\s,\varphi)=\sum_{b\in M_r}(U(\s_b)-U(\varphi_b)).$$
\vskip 0.4 truecm

{\bf Definition 3.1.} {\it Let $GS$ be the complete set of all
ground states of the relative Hamiltonian $H$. A ball $b\in M_r$ is
said to be an {\sl improper} ball of the configuration $\s$ if $\s_b
\ne \f_b$ for any $\f\in GS.$ The union of the improper balls of a
configuration $\s$ is called the {\sl boundary} of the configuration
and denoted by} $\d(\s).$

\vskip 0.4 truecm

{\bf Definition 3.2.} {\it The relative Hamiltonian $H$ with the set
of ground states $GS$  satisfies the Peierls condition if for any
$\f\in GS$ and any configuration $\s$ coinciding almost everywhere
with $\f$ (i.e. $|\{x\in V: \s(x)\ne\f(x)\}|<\infty$)
$$H(\s,\f)\geq \lambda |\d(\s)|,$$
where $\l$ is a positive constant which does not depend on $\s$, and
$|\d(\s)|$ is the number of  balls in} $\d(\s).$

\vskip 0.4 truecm

{\bf Theorem 3.3.} {\it If assumptions A1-A3 are satisfied
 then the Peierls condition holds.}

\vskip 0.5 truecm

{\it Proof.} Suppose $\s$ coincides almost everywhere with  a ground
state $\f\in GS$ then we have $U(\s_b)- U^{\min}\geq \l_0$ for any
$b\in \d(\s)$ since $\f$ is a ground state. Thus
$$H(\s,\f)=\sum_{b\in M_r}(U(\s_b)-U(\f_b))=
\sum_{b\in \d(\s)}(U(\s_b)-U^{\min})\geq \l_0|\d(\s)|.$$ Therefore,
the Peierls condition is satisfied for $\lambda=\l_0$. The theorem
is proved.

\vskip 0.4 truecm

\section{Contours on Cayley tree}

Let $\L\subset V$ be a finite set. Let $\s^{(i)}_{\L^c}\equiv i$,
$i=1,...,s$ be a constant configuration outside of $\L .$ For each
$i$ we extend the configuration $\s_\L$ inside $\L$ to the entire
tree by the $i$th constant configuration and denote it by
$\s^{(i)}_\L$. The set of such configurations we denote by
$\O^{(i)}_\L.$

Now we are going to recall a construction of the subcontours (see
[20]). Note that our definition (see definition 4.3 below) of a
contour depends on $r$, at $r=1$ we get a contour defined in [20].
But the definition of a subcontour does not depend on $r.$

Consider $V_n$ and for a given configuration $\s^{(i)}_\L\in
\O^{(i)}_\L$ denote $V^{(j)}_n\equiv V^{(j)}_n(\s^{(i)}_\L)=\{t\in
V_n: \s^{(i)}_\L(t)= j\}, j=1,...,q, j\ne i.$ Let
$G^{n,j}=(V^{(j)}_n, L^{(j)}_n) $ be a graph such that
$$L^{(j)}_n=\{l=<x,y>\in L: x,y\in V^{(j)}_n\},\ \  j=1,...,q, j\ne i.$$
It is clear, that for a fixed $n$ the graph $G^{n,j}$ contains a
finite number $(=m)$ of maximal  connected subgraphs $G^{n,j}_p$
i.e.
$$G^{n,j}=\{G^{n,j}_1, ..., G^{n,j}_m\}, \ \ G^{n,j}_p=(V^{(j)}_{n,p},
L^{(j)}_{n,p}), \ \ p=1,...,m; j\ne i.$$ Here $V^{(j)}_{n,p}$ and
$L^{(j)}_{n,p}$ are the set of vertexes and  edges of $G^{n,j}_p$,
respectively.

Two edges $l_1,l_2\in L, \ \ (l_1\ne l_2)$ are called {\it nearest
neighboring edges} if $|i(l_1)\cap i(l_2)|=1$, and we write $<l_1,
l_2>_1.$

For any subgraph $K\subset \G^k$ denote by $E(K)$ the set of
edges, by $V(K)$ the set of vertices of $K$ and
$$B(K)=\{l\in L\setminus E(K): \exists l_1\in E(K) \ \ \mbox{such that}
\ \ <l,l_1>_1\}. $$

{\bf Definition 4.1.} {\it  An edge $l=<x,y>\in L_{n+1}$ is called a
{\it boundary edge} of the configuration $\s^{(i)}_{V_n}$ if}
 $\s^{(i)}_{V_n}(x)\ne \s^{(i)}_{V_n}(y) .$
\vskip 0.4 truecm

 The set of boundary edges of $\s^{(i)}_{V_n}$ is
called {\it edge boundary} $\d_1 (\s^{(i)}_{V_n})\equiv \d_1$ of
the configuration.

The (finite) set $B(G^{n,j}_p), j=1,...,q, j\ne i, p=1,...,m $
(together with a given configuration on it) is called {\it
subcontour} of the boundary $\d_1.$

The set $V^{(j)}_{n,p}, \ \ j=1,...,q, j\ne i, p=1,...,m$ is
called {\it interior}  of  $B(G^{n,j}_p)$, and is denoted by
Int$B(G^{n,j}_p)$. The set of edges from a subcontour $T$ is
denoted by supp$T$ . The configuration $\s^{(i)}_{V_n}$ takes the
same value $j$ at all points of the connected component
$G^{n,j}_p$. This value $v= v(G^{n,j}_p)$ is called the {\it mark}
of the subcontour and denoted by $v(T)$, where $T=B(G^{n,j}_p).$

The collection of subcontours $\tau=\tau(\s^{(i)}_{V_n})=\{T_p\}$
generated by the edge boundary $\d_1=\d_1(\s^{(i)}_{V_n})$ of
$\s^{(i)}_{V_n}$ has the following properties

(a) Every subcontour $T\in \tau$ lies inside the set $V_{n+1}.$

(b) For every two subcontours $T_1, T_2\in \tau$ their supports
supp$T_1$ and supp$T_2$ satisfy $|{\rm supp} T_1 \cap {\rm supp}T_2
|\in \{0, 1\}.$

(c) For any two  subcontours $T_1, T_2\in \tau$ with $|$supp$T_1
\cap $supp$T_2|= 1$ we have $v(T_1)\ne v(T_2).$

The distance ${\rm dist}(T_1,T_2)$ is defined by
$${\rm dist}(T_1,T_2)=\min_{x\in V(T_1)\atop y\in V(T_2)}d(x,y),$$
where $d(x,y)$ is the distance between $x,y\in V$ (see section 2.1).

Recall $r'=[{r+1\over 2}].$

\vskip 0.4 truecm

{\bf Definition 4.2.} {\it The subcontours $T_1,T_2$ are called
{\sl adjacent} if ${\rm dist}(T_1,T_2)\leq 2(r'-1).$ A set of
subcontours ${\cal A}$ is called {\sl connected} if for any two
subcontours $T_1,T_2\in {\cal A}$ there is a collection of
subcontours $T_1={\tilde T}_1, {\tilde T}_2,...,{\tilde T}_l=T_2$
in ${\cal A}$ such that for each $i=1,...,l-1$ the subcontours
${\tilde T}_i$ and ${\tilde T}_{i+1}$ are adjacent.}

\vskip 0.4 truecm

{\bf Definition 4.3.} {\it Any maximal connected set (component) of
subcontours is called {\sl contour} of the set} $\d_1.$

\vskip 0.4 truecm

For contour  $\g=\{T_p\}$ denote  ${\rm Int}\g=\cup_p{\rm Int}T_p.$

\vskip 0.4 truecm

{\it Remarks.} 1. Note that Definition 4.3 of contours coincides
with the Definition 2 of [20] for $r=1$. But Definition 4.3 is
better than corresponding Definition 11 of [21] for $r=2$. Because,
for $r=2$ from the definition 4.2 we have ${\rm dist}(T_1, T_2)=0$
i.e. the subcontours do not interact if the distance between them is
$\geq 1$ but in [21] the condition was like ${\rm dist}(T_1,T_2)\leq
2.$

2. Our definition of a contour is slightly different from the
definition of contour of Hamiltonians on $Z^d,$ $d\geq 2$ (see [17],
[22]). For any two contours $\g, \g'$ we have ${\rm
dist}(\g,\g')>2(r'-1)$. Thus our contours do not interact. This
means that for any $\s\in \O$ there is no a ball $b\in \d(\s)$ with
$b\cap \g\ne\emptyset$ and $b\cap \g'\ne \emptyset.$ Such property
allows as to use a {\it contour-removal} operation. This operation
is similar to the one in ordinary Peierls argument [7]: Given a
family of contours defining a configuration $\s\in \O^{(i)}_\L,$ the
family obtained by omitting one of them is also the family of
contours of a (different) configuration in $\O^{(i)}_\L.$ There is
an algorithm of the contour-removal operation to obtain a new
configuration as follows. Take the configuration $\s$ and change all
the spins in the interior of $\g$ (which must be removed) to value
$i.$ This makes $\g$ disappear, but leaves intact the other
contours.

 \vskip 0.4 truecm

For a given (sub)contour $\g$ denote
$${\rm imp}\g=\{b\in \d: b\cap \g\ne
\emptyset\}, \ \ |\g|=|{\rm imp}\g|.$$

By the construction we have ${\rm imp}\g\cap {\rm imp}\g'=\emptyset$
for any contours $\g\ne\g'.$

For $A\subset V$ denote
$$C(A)=\{b\in M_r : b\cap A\ne \emptyset\};$$
$$D(A)=\{x\in V\setminus A : \exists y\in A, \ \
{\rm such\ \  that} \ \ <x,y>\}.$$
 \vskip 0.4 truecm

{\bf Lemma 4.4.}  {\it Let $K$ be a connected subgraph of the Cayley
tree $\G^k$ of order $k\geq 2$, such that $|V(K)|=n,$ then}

(i). $|D(V(K))|=(k-1)n+2.$

(ii). $|C(V(K))|=k^{r'-1}((k-1)n+2).$

\vskip 0.4 truecm

{\it Proof.} (i). We shall use the induction over $n .$ For $n=1$
and 2 the assertion is trivial. Assume for $n=m$ the lemma is true
i.e. from $|V(K)|=m$ follows that $|D(V(K))|=(k-1)m+2.$ We shall
prove the assertion for $n=m+1$ i.e. for $\tilde{K}=K\cup \{x\}.$
Since $\tilde{K}$ is connected graph we have $x\in D(V(K))$ and
there is a unique $y\in S_1(x)=\{u\in V: d(x,u)=1\}$ such that
$y\in V(K).$ Thus $D(V(\tilde{K}))=(D(V(K))\setminus \{x\})\cup
(S_1(x)\setminus \{y\}).$ Consequently,
$$|D(V(\tilde{K}))|=|D(V(K))|-1+k=(k-1)(m+1)+2.$$

(ii). Using (i) we obtain $|C(V(K))|=u_{r'}$, where $u_{r'}$ is
the last term of the collection $u_1,u_2,...,u_{r'}$ which is
defined by the following recurrent relations
$$u_l=2+(k-1)\sum_{i=0}^{l-1}u_i, \ \ l=1,2,...,r',\ \  u_0=n.\eqno (4.1) $$
Iterating (4.1) we get $u_1=(k-1)n+2$, $u_2=k((k-1)n+2),$ then
using the induction over $l$ we obtain $u_l=k^{l-1}((k-1)n+2).$
This completes the proof.

Let us define a graph structure on $M_r$ as follows. Two balls $b,
b'\in M_r$ are connected by an edge if their centers are nearest
neighbors. Denote this graph by $G(M_r).$ Note that the graph
$G(M_r)$ is a Cayley tree of order $k\geq 1.$ Here the vertices of
this graph are balls of $M_r.$ Thus Lemma 1.2 of [5] can be
reformulated as follows

\vskip 0.4 truecm

{\bf Lemma 4.5.} {\it Let ${\tilde N}_{n,G}(x)$ be the number of
connected subgraphs $G'\subset G(M_r)$ with $x\in V(G')$ and
$|V(G')|=n.$ Then}
$${\tilde N}_{n,G}(x)\leq (e k)^n.$$

 For
$x\in V$ we will write $x\in \g$ if $x\in V(\g).$

Denote $N_l(x)=|\{\g: x\in \g, |\g|=l\}|,$ where as before
$|\g|=|{\rm imp}\g|.$

\vskip 0.4 truecm

{\bf Lemma 4.6.} {\it If $k\geq 2$ then}
$$N_l(x)\leq C_0\t^l, \eqno(4.2)$$
where $C_0=1+{k+1\over k-1}(k^{r'}-1),$
$\t=\t(k,r)=(2ek)^{2(k+1)(r'-1)k^{r'-1}+2}$.

\vskip 0.4 truecm

{\bf Proof.} Denote by $K_\g$ the minimal connected subgraph of
$\G^k,$ which contains a contour $\g=\{\g_1,...,\g_m\}, m\geq 1, $
where $\g_i $ is subcontour. Put
$$M^-_{r,\g}=\{x\in {\rm Int}\g: {\rm dist}(x,V\setminus {\rm
Int}\g)> r'\};$$
$$M^0_{r,\g}=\{x\in {\rm Int}\g: {\rm dist}(x,V\setminus {\rm
Int}\g)\leq r'\};$$
$$M^+_{r,\g}=\{x\in V\setminus {\rm Int}\g: {\rm dist}(x,{\rm
Int}\g)\leq r'\};$$
$$Y_\g=V(K_\g)\setminus \big({\rm Int}\g\cup D({\rm Int}\g)\big).$$
We have
$$|\g|=|M^0_{r,\g}|+|M^+_{r,\g}|;$$
$$|C(V(K_\g))|\leq |M^-_{r,\g}|+|\g|+|C(Y_\g)|.\eqno(4.3)$$

For any $k\geq 2$, $r\geq 1$ by Lemma 4.4 we have
$$|M^-_{r,\g}|={|D(M^-_{r,\g})|-2\over k-1}<|D(M^-_{r,\g})|\leq
|M^0_{r,\g}|<|\g|.\eqno (4.4)$$ Note that $0\leq |Y_\g|\leq
2(m-1)(r'-1).$ Thus
$$|C(T_\g)|\leq 2(m-1)(r'-1)|C(\{y\})|, \eqno(4.5)$$
where $y$ is an arbitrary point of $Y_\g.$ By Lemma 4.4. we have
$|C(\{y\})|=k^{r'-1}(k+1)$ since $|V(\{y\})|=1.$ Hence from
(4.3)-(4.5) we get
$$|C(V(K_\g))|<2|\g|+2(m-1)(r'-1)(k+1)k^{r'-1}. \eqno(4.6)$$
Since $\g$ contains $m$ subcontours we have $m< |\g|.$  A
combinatorial calculations show that
$$N_l(x)\leq C_0\sum_{m=1}^l{2l+2k^{r'-1}(k+1)(r'-1)(m-1)\choose l}{\tilde
N}_{2l+2k^{r'-1}(k+1)(r'-1)(m-1), \G^k}(b_x), \eqno (4.7)$$ where
${\tilde N}_{l, \G^k}$ is defined in Lemma 4.5 and $b_x$ is a ball
$b\in M_r$ such that $x\in b$. Using inequality ${n\choose l}\leq
2^{n-1}, l\leq n$ and Lemma 4.5 from (4.7) we get (4.2). The lemma
is proved.

\section{ Non-uniqueness of Gibbs measure}

For $\s_n\in \O^{(i)}_{V_n}$ the conditional Hamiltonian (2.3) has
the form

$$H(\s_n)\equiv
H(\s_n\big | \s_{V^c_n}=i)=\sum_{b\in M_r:\atop b\cap
V_n\ne\emptyset}U(\s_{n,b})=$$
$$\sum_{b\in \d(\s_n)}(U(\s_{n,b})-U^{\min})+|C(V_n)|U^{\min},
\eqno(5.1)$$ where $\s_{n,b}=(\s_n)_b.$

 The Gibbs measure on the
space $\O^{(i)}_{V_n}$ with boundary condition $\s^{(i)}$ is defined
as
$$\mu^{i}_{n,\b}(\s_n)={\bf Z}_{n,i}^{-1}\exp(-\b H(\s_n)), \eqno(5.2)$$
where ${\bf Z}_{n,i}$ is the normalizing factor.

 Let us consider a sequence of balls on $\G^k$
$$V_1\subset V_2\subset ... \subset V_n\subset ..., \ \ \cup V_n=V,$$
and $s$ sequences of boundary conditions outside these balls:
$$\s^{(i)}_n\equiv i, n=1,2,..., i=1,...,s .$$
By very similar argument of proof of the lemma 9.2 in [14] one can
prove that each of $s$ sequences of measures $\{\m^{i}_{n,\b},
n=1,2,...\}, i=1,...,s$ contains a convergent subsequence.

We denote the corresponding limits by $\m^i_{\b}, i=1,...,s$.
 Our purpose is to show for
 a sufficiently large $\b$ these measures are different.
\vskip 0.4 truecm

{\bf Lemma 5.1.} {\it Suppose assumptions A1-A3 are satisfied. Let
$\g$ be a fixed contour and $p_i(\g)=\mu^i_\b(\s_n: \g\in \d(\s_n)).
$ Then
$$p_i(\g)\leq \exp\{-\b\l_0|\g|\}, \eqno(5.3)$$
where $\l_0$ is defined by formula (3.1).}
\vskip 0.4 truecm

{\it Proof.} Put $\O_\g=\{\s_n\in \O^{(i)}_{V_n}: \g\subset
\d(\s_n)\}$, $\O_\g^0=\{\s_n: \g\cap \d=\emptyset\}$  and define a
(contour-removal)  map $\chi_\g:\O_\g \to \O_\g^0$ by

$$\chi_\g(\s_n)(x)=\left\{\begin{array}{ll}
i & \textrm{if \ \ $ x\in {\rm Int}\g$}\\
\s_n(x)& \textrm{if \ \ $x\notin {\rm Int}\g.$} \\
\end{array}\right.
$$

For a given $\g$ the map $\chi_\g$ is one-to-one map.
 For any $\s_n\in \O^{(i)}_{V_n}$ we have
$$|\d(\s_n)|=|\d(\chi_\g(\s_n))|+|\g|.$$
Consequently, using (5.1) one finds
$$p_i(\g)={\sum_{\s_n\in \O_\g}\exp\{-\b \sum_{b\in \d(\s_n)}(U(\s_{n,b})-U^{\min})\}\over
\sum_{{\tilde\s}_n}\exp\{-\b \sum_{b\in
\d({\tilde\s}_n)}(U({\tilde\s}_{n,b})-U^{\min})\}}\leq
$$
$${\sum_{\s_n\in \O_\g}\exp\{-\b \sum_{b\in
\d(\s_n)}(U(\s_{n,b})-U^{\min})\}\over \sum_{{\tilde\s}_n\in
\O^0_\g}\exp\{-\b \sum_{b\in
\d({\tilde\s}_n)}(U({\tilde\s}_{n,b})-U^{\min})\}}=$$
$${\sum_{\s_n\in \O_\g}\exp\{-\b \sum_{b\in
\d(\s_n)}(U(\s_{n,b})-U^{\min})\}\over \sum_{{\tilde\s}_n\in
\O_\g}\exp\{-\b \sum_{b\in
\d(\chi_\g({\tilde\s}_n))}(U(\chi_\g({\tilde\s}_{n,b}))-U^{\min})\}}.
\eqno(5.4)$$ Since $\s_{n,b}=\chi_\g(\s_{n,b}),$ for any $b\in
\d(\s_n)\setminus {\rm imp}\g$ we have
$$\sum_{b\in \d(\s_n)}(U(\s_{n,b})-U^{\min})=S_1+S_2, \eqno(5.5)$$ where
$S_1=\sum_{b\in \d(\chi_\g(\s_n))}(U(\s_{n,b})-U^{\min});$
$S_2=\sum_{b\in {\rm imp}\g}(U(\s_{n,b})-U^{\min}).$

Note that for a fixed $\g$ the sum $S_2$ does not depend on
configuration $\s_n\in \O_\g.$ Indeed, by our construction $\g$ is
a contour of $\d(\s_n)$ iff $\s_n(x)=i$ for any $x\in M^+_{r,\g}.$
Consequently, ${\rm imp}\g$ and  $S_2$ do not depend on
$\s_n\in\O_\g.$

Hence, (5.4) implies that

$$p_i(\g)\leq \exp(-\b S_2). \eqno(5.6)$$

By assumptions A1-A3 we have $U(\s_{n,b})-U^{\min}\geq \l_0>0,$ for
any $b\in {\rm imp}\g.$ Thus from (5.6) one gets (5.3). The lemma is
proved.

Now using Lemmas 4.6 and 5.1 by very similar argument of [20] one
can prove the following
 \vskip 0.4 truecm

{\bf Lemma 5.2.} {\it If assumptions A1-A3 are satisfied then for
fixed $x\in \L$ uniformly by $\L$ the following relation holds}
$$\mu^i_\b(\s_\L: \s(x)=j)\to 0, j\ne i \ \ as \ \ \b\to \infty.$$

This lemma implies the main result, i.e.
\vskip 0.4 truecm

{\bf Theorem 5.3.} {\it If A1-A3 are satisfied then for all
sufficiently large $\b$ there are at least $s$ (=number of ground
states) Gibbs measures for the model (2.2) on Cayley tree of order
$k\geq 2.$}

\section{Examples} In this section we shall give several examples
with the properties A1-A3.
\vskip 0.3 truecm

{\it 6.1. $q-$ component models.} Note that under some suitable
conditions on the parameters of $q$-component models (with nearest
neighbor interactions) on Cayley tree (see [20]) the assumptions
A1-A3 are satisfied. In particular, the {\it ferromagnetic}  Ising,
Potts and  SOS models have the properties A1-A3.

 \vskip 0.3 truecm

{\it 6.2. The Potts model with competing interactions $(k=2, q=3)$.}
Consider the Hamiltonian
$$H (\sigma) =J_1\sum\limits _ {\langle x, y
\rangle, \atop x, y \in V} \delta _ {\sigma (x) \sigma
(y)}+J_2\sum\limits _ {x, y \in V: \atop d (x, y) =2}\delta _
{\sigma (x) \sigma (y)}, \eqno (6.1)
$$ where $J = (J_1, J_2) \in R^2$, $\s(x)\in \Phi=\{1,2,...,q\}$ and
$\delta$ is the Kronecker's symbol i.e.
$$\delta_{u, v}=\left\{\begin{array}{ll}
1 & \textrm{if \ \
$ u=v$}\\
0 & \textrm{if \ \ $u\ne v.$}\\
\end{array}\right.
$$

Note that the Ising model with competing interactions (see [21]) is
a particular case of the model (6.1). For the model (6.1) with
$k=2$, $q=3$ we put
$$U (\sigma_b) \equiv U (\sigma_b, J) = \frac {1} {2} J_1\sum\limits _ {\langle x, y
\rangle, \atop x, y \in b} \delta _ {\sigma (x) \sigma (y)}
+J_2\sum\limits _ {x, y \in b:\atop d (x, y) =2} \delta _ {\sigma
(x) \sigma (y)}. \eqno (6.2) $$ A simple calculations show that
$${\bf U}=\{U(\s_b)\}= \big\{\frac{3}{2}J_1+3J_2,\ \ J_1+J_2,\ \
 3J_2,\ \ \frac {1}{2}J_1,\ \  J_2,\ \  \frac {1}{2} J_1+J_2\big\}. $$
 By similar argument of [21] (pages 221-223) one can show that
 for the model (6.1) the assumptions A1-A3 are satisfied if $J\in \{J\in
 R^2: J_1<0, J_1+4J_2<0\}.$

\vskip 0.3 truecm

 {\it 6.3. A model with the interaction radius $r\geq 1$.} For $A\subset V$ let us define a
 generalized Kronecker symbol as the function $U_0(\s_A):\O_A\to
 \big\{|A|-1, |A|-2,..., |A|-\min\{|A|, |\Phi|\}\big\}$ by
 $$ U_0(\s_A)=|A|-|\s_A\cap \Phi|, \eqno(6.3)$$
 Note that if $|A|=2,$ say,  $A=\{x,y\},$ then $
 U_0(\{\s(x),\s(y)\})=\delta_{\s(x),\s(y)}.$
 Now consider the Hamiltonian
 $$H(\s)=-J\sum_{b\in M_r}U_0(\s_b), \eqno(6.4)$$
 where $J\in R.$

 It is easy to see that if $J>0$ then the assumptions A1-A3
  are satisfied for any $r\geq 1$ and $k\geq 2.$

 \vskip 0.3 truecm

 {\bf Acknowledgments.} The work supported by NATO
Reintegration Grant : FEL. RIG. 980771. A part of this work was
done within the scheme of Junior Associate at the ICTP, Trieste,
Italy and the author thanks ICTP for providing financial support
and all facilities (in May - August 2006). The final part of this
work was done at  the IHES, Bures-sur-Yvette, France. I thank the
IHES for support and kind hospitality (in October - December
2006). I also gratitude to professors M. Cassandro, M.
Kontsevitch, Yu.Suhov and F.Mukhamedov for many helpful
discussions.

\vskip 0.3 truecm

{\bf References}

1. Baxter, R.J.:  Exactly  Solved Models in Statistical Mechanics,
 London/New York: Academic Press, 1982.

2. Biskup, M., Borgs, C., Chayes, J. T., Koteck\'y, R.: Partition
function zeros at first-order phase transitions: Pirogov-Sinai
theory. J. Stat. Phys. {\bf 116}, 97-155 (2004)

3. Bleher, P.M.,  Ruiz,J., Schonmann, R.H.,  Shlosman, S.,
Zagrebnov, V.A.: Rigidity of the critical phases on a Cayley tree.
Moscow Math. J. {\bf 3}, 345-362 (2001)

4. Bleher, P.M., Ganikhodjaev, N.N.: On pure phases of the Ising
model on the Bethe lattice.  Theor. Probab. Appl. {\bf 35}, 216-227
(1990)

5. Borgs, C.: Statistical physics expansion methods in combinatorics
and computer science,
http://research.microsfort.com/~borgs/CBMS.pdf, 2004.

6. Bovier, A., Merola, I., Presutti, E., Zahradnik, M.: On the Gibbs
phase rule in the Pirogov-Sinai regime. J. Stat. Phys. {\bf 114},
1235-1267 (2004)

7. Fern$\acute{\textrm {a}}$ndez, R.: Contour ensembles and the
description of Gibbsian probability distributions at low
temperature. www.univ-rouen.fr/LMRS/persopage/Fernandez, 1998.

8  Ganikhodjaev, N.N., Rozikov, U.A. A description of periodic
extremal Gibbs measures of some lattice models on the Cayley tree.
Theor. Math. Phys. {\bf 111}, 480-486 (1997)

9. Ganikhodjaev, N. N., Rozikov, U. A. The Potts model with
countable set of spin values on a Cayley tree.  Lett. Math. Phys.
{\bf 75}, 99-109 (2006)

10. Georgii, H.-O.: Gibbs measures and phase transitions, Berlin:
Walter de Gruyter, 1988.

11. Grimmett, G.: The random-cluster model, Berlin: Springer, 2006.

12. Lebowitz, J. L., Mazel, A. E.: On the uniqueness of Gibbs states
in the Pirogov-Sinai theory. Commun. Math. Phys. {\bf 189}, 311-321
(1997)

13.  Martin, J.B., Rozikov, U.A., Suhov, Yu.M.: A three state
hard-core model on a Cayley tree.  J. Nonlinear Math. Phys. {\bf
12}, 432-448 (2005)

14.  Minlos, R.A.:  Introduction to mathematical statistical
physics, University lecture series,  v.19, AMS,  2000.

15. Mukhamedov, F.M.,  Rozikov, U.A.: On Gibbs measures of models
with competing ternary and binary interactions and corresponding von
Neumann algebras. I, II.  J. Stat. Phys. {\bf 114}, 825-848 (2004);
{\bf 119}, 427-446 (2005)

16. Peierls, R.: On Ising model of ferro magnetism.  Proc. Cambridge
Phil. Soc. {\bf 32}, 477-481 (1936).

17. Pirogov, S.A., Sinai,Ya. G.: Phase diagrams of classical lattice
systems.I, II.  Theor. Math. Phys. {\bf 25}, 1185-1192 (1975); {\bf
26}, 39-49 (1976)

18. Rozikov, U. A., Suhov, Yu.M.: A hard-core model on a Cayley
tree: an example of a loss network.  Queueing Syst. {\bf 46},
197-212 (2004)

19. Rozikov, U.A.: An example of one-dimensional phase transition.
Siber. Adv. Math. {\bf 16}, 121-125 (2006)

20. Rozikov, U.A.: On $q-$ component models on Cayley tree: contour
method.  Lett. Math. Phys. {\bf 71}, 27-38 (2005)

21. Rozikov, U. A.: A constructive description of ground states and
Gibbs measures for Ising model with two-step interactions on Cayley
tree.  J. Stat. Phys. {\bf 122}, 217-235 (2006)

22. Sinai, Ya.G.:  Theory of phase transitions: Rigorous Results,
Oxford: Pergamon, 1982.

23. Zachary, S.: Countable state space Markov random fields and
Markov chains on trees.  Ann. Prob. {\bf 11}, 894-903 (1983)

24. Zahradnik, M.: An alternate version of Pirogov-Sinai theory.
Commun. Math. Phys. {\bf 93}, 559-581 (1984)

25. Zahradnik, M.: A short course on the Pirogov-Sinai theory.
Rendiconti Math. Serie VII. {\bf 18}, 411-486 (1998)

\end{document}